\documentclass{llncs}
\pdfoutput=1
\usepackage{makeidx}  % allows for indexgeneration

\usepackage[pdftex]{graphicx}
\usepackage[pdftex]{color}
\usepackage{url}
\usepackage{comment}
\usepackage{multirow}

\begin{document}
\frontmatter          % for the preliminaries
\pagestyle{headings}  % switches on printing of running heads
\mainmatter              % start of the contributions

\def\etal{et~al.\,}
\hyphenation{Nationaldemokr-atische}

\pdfinfo{
/Title (An Analysis of Interactions Within and Between Extreme Right Communities in Social Media)
/Author (Derek O'Callaghan, Derek Greene, Maura Conway, Joe Carthy, Padraig Cunningham)
/Subject (network analysis, social media, community detection, topic modelling, Twitter, extreme right)
/Keywords (network analysis, social media, community detection, topic modelling, Twitter, extreme right) }

\title{An Analysis of Interactions Within and Between Extreme Right Communities in Social Media}
\titlerunning{TBD}  % abbreviated title (for running head)
\author{Derek O'Callaghan\inst{1}, Derek Greene\inst{1}, Maura Conway\inst{2}, Joe Carthy\inst{1}, P\'{a}draig Cunningham\inst{1}}
\authorrunning{Derek O'Callaghan \etal} % abbreviated author list (for running
% head)
\tocauthor{Derek O'Callaghan, Derek Greene, Maura Conway, Joe Carthy, P\'{a}draig Cunningham}
\institute{School of Computer Science \& Informatics, University College Dublin,\\
\email{\{derek.ocallaghan,derek.greene,joe.carthy,padraig.cunningham\}@ucd.ie}
\and
School of Law \& Government, Dublin City University,\\
\email{maura.conway@dcu.ie}
}

\maketitle              % typeset the title of the contribution

\begin{abstract}
\begin{comment}

The abstract should summarize the contents of the paper
using at least 70 and at most 150 words. It will be set in 9-point
font size and be inset 1.0 cm from the right and left margins.
There will be two blank lines before and after the Abstract. \dots
\end{comment}

Many extreme right groups have had an online presence for some time through the use of dedicated websites. This has been accompanied by
increased activity in social media platforms in recent years, enabling the dissemination of extreme right content to a wider audience. In
this paper, we present an analysis of the activity of a selection of such groups on Twitter, using network representations based on
reciprocal follower and interaction relationships, while also analyzing topics found in their corresponding tweets. International
relationships between certain extreme right groups across geopolitical boundaries are initially identified. Furthermore, we also discover
stable communities of accounts within local interaction networks, in addition to associated topics, where the underlying extreme right
ideology of these communities is often identifiable.

\keywords{network analysis, social media, community detection, topic modelling, Twitter, extreme right}
\end{abstract}

\section{Introduction}

Groups associated with the extreme right have maintained an online presence for some time \cite{hoffman1996web,Burris2000}, where dedicated
websites have been employed for the purposes of content dissemination and member recruitment. Recent years have seen increased activity by
these groups in social media platforms, given the potential to access a far wider audience than was previously
possible~\cite{BartlettDigitalPopulism2011,ZwischenPropaganda2011}. In this paper, we present an analysis of the activity of a selection of
these groups on Twitter, where the focus is upon groups of a fascist, racist, supremacist, extreme nationalist or neo-Nazi nature, or some
combination of these. Twitter's features enable extreme right groups to disseminate hate content with relative ease, while also facilitating
the formation of communities of accounts around variants of extreme right ideology. Message posts (\textit{tweets}) by members of these
groups, to which access is usually unrestricted, are often used to redirect users to content hosted on external websites, including
dedicated websites managed by particular groups, or content sharing platforms such as YouTube.

For the purpose of this analysis, we have retrieved data for a selection of identified extreme right Twitter accounts from eight countries.
Our initial objective is the identification of international relationships between certain groups that transcend geopolitical boundaries.
This involves the analysis of two network representations of the accounts from the eight country sets, based on reciprocal follower and
interaction relationships. Here, interactions are derived from observations of mentions and retweets between accounts. It appears that a
certain amount of international awareness exists between accounts based on the follower relationship, while interactions indicate stronger
relationships where linguistic and geographical proximity are highly influential.

This leads to our next objective of analyzing communities of extreme right accounts found within local interaction networks, where locality
is considered in terms of nationality or linguistic proximity. Tweet content is also analyzed for the purpose of generating interpretable
descriptions for the detected communities, in addition to the discovery of latent topics associated with interactions between the member
accounts. We find that matrix factorization techniques are more suitable for topic analysis of these particular data sets. By using the same
account profile document representation for both community description generation and topic discovery, it is possible to generate a mapping
between the detected communities and their associated topics. Each community description and corresponding topic mapping can then be used in
conjunction with manual analysis of the account profiles, tweets and external websites to provide an interpretation of the underlying
community ideology. While we observe some community division along electoral and non-electoral lines, this is not clear in all cases. Other
notable findings include communities of a more traditional conservative nature, opposition to bodies such as the EU, and the influence of
concerns such as counter-Jihad on international relationships.

In Section~\ref{relatedwork}, we provide a description of related work based on the online activities of extremist groups. The collection of
the Twitter data sets is then discussed in Section~\ref{curation}. Analysis of the international relationships between extreme right groups
from the eight countries is presented in Section~\ref{global}. Next, in Section~\ref{local}, we describe the discovery of extreme right
communities within local interaction networks, including the methodology used for network derivation, community detection, stability
ranking, description generation and topic analysis. We focus on two case studies using English and German language networks, where we offer
an interpretation of a selection of these communities. Finally, the overall conclusions are discussed in Section~\ref{conclusions}, and
some suggestions for future work are made.

\section{Related Work}
\label{relatedwork}

The online activities of different varieties of extremist groups including those associated with the extreme right have been the subject of
a number of studies. Burris~\etal used social network analysis to study a network based on the links between a selection of white
supremacist websites~\cite{Burris2000}. They found this network to be relatively decentralized with multiple centres of influence, while
also appearing to be mostly undivided along doctrinal lines. Similar decentralization and multiple communities were found by Chau and Xu in
their study of networks built from users contributing to hate group and racist blogs~\cite{Chau:2007:MCR:1222244.1222622}. They also found
that some of these groups exhibited transnational characteristics. In a similar approach to that of Burris~\etal, Tateo analyzed groups
associated with the Italian extreme right, using networks based on links between group websites~\cite{JCC4:JCC410}. Caiani and Wagemann
studied similar Italian groups along with those from the German extreme right, where they found the German network to be structurally
centralized to a greater extent than that of the Italian groups~\cite{doi:10.1080/13691180802158482}. The contents of websites belonging to
central nodes within Russian extreme right networks were analyzed by Zuev \cite{NANA:NANA430}. In their review of the conservative movement
in the USA, Blee and Creasap~\cite{Blee2010} discuss the engagement in online activity as part of an overall mobilization strategy by the
more extremist groups within it. 

The potential for online radicalization through exposure to jihadi video content on YouTube was investigated by Bermingham~\etal, where it
was suggested that a potentially increased online leadership role may be attributed to users claiming to be women, according to centrality,
network density and average speed of communication~\cite{Bermingham:2009:CSN:1602240.1602692}. Sureka~\etal also studied the activity of
extremist users within YouTube, investigating content properties along with hidden network communities~\cite{Sureka_Kumaraguru_2010}.
Bartlett~\etal performed a survey of European populist party and group supporters on Facebook~\cite{BartlettDigitalPopulism2011}, while
Baldauf~\etal also investigated the use of Facebook by the German extreme right~\cite{ZwischenPropaganda2011}. As the majority of this work
involved the study of dedicated websites managed by extreme right groups, we believed that an analysis of their activity in social media,
focusing specifically on extreme right Twitter communities, would complement this by providing additional insight into the overall online
presence of these groups.

\section{Data}
\label{curation}

Twitter data was collected to facilitate the analysis of contemporary online extreme right activity. We identified initial sets of relevant
accounts for a selection of countries, where the country selection was informed by prior knowledge of extreme right groups. Based on
earlier studies~\cite{ZwischenPropaganda2011,doi:10.1080/13691180802158482}, the criteria used to identify relevant accounts included
profiles containing references to known groups or employing extreme right symbols; recent tweet activity; similar Facebook/YouTube accounts;
reciprocal follower relationships with known relevant accounts; accounts with self-curated Twitter lists containing relevant accounts; extreme right
media accounts such as record labels and concert organisers. Details of these country sets can be found in Table~\ref{tab:coresets}.

\begin{table}[!t]
\caption{Data set sizes for eight countries of interest.}
\begin{center}
\begin{tabular}{| l | c |}
\hline 
\emph{Country} & \emph{Number of Accounts}
\\
\hline \hline 
France \hspace{4em} & 25  \\ \hline
Germany & 53  \\ \hline
Greece & 45  \\ \hline
Italy & 17  \\ \hline
Spain & 43  \\ \hline
Sweden & 21  \\ \hline
UK & 32  \\ \hline
USA & 32  \\ \hline
\end{tabular}
\end{center}
\label{tab:coresets}
\end{table}

Certain accounts were not included, such as inactive accounts, or those that were not deemed to be related to the extreme right. These included
traditional conservative (e.g. centre-right) accounts, non-conformists/anti-establishment accounts considered to be left-wing, and conspiracy
theorists. As we were initially interested in non-electoral extreme right groups~\cite{GoodwinRadicalRight2012}, higher-profile politicians
or political parties were ignored for the most part, with a minor number of these accounts included where it was felt that there was a close
association with relevant accounts. An obstacle was the language barrier, where the use of online translation tools did not always prove 
 helpful in the interpretation of ambiguous profiles. In cases where the relevance of an account profile was inconclusive, that account was
ignored. Twitter data including followers, friends, tweets and list memberships were retrieved for each of the selected accounts during the
period March -- August 2012, as limited by the Twitter API restrictions effective at the time.

\section{International Relationships}
\label{global}

We began with an analysis of the international relationships between the identified extreme right groups, based on interactions between the
accounts from the eight country sets. An \emph{interaction} is defined as one account ``mentioning'' another account within a tweet, or an
account ``retweeting'' a tweet generated by another account. Both types of event were included in order to address issues of data sparsity
and incompleteness. We were particularly interested in reciprocal activity between accounts, where such activity can potentially indicate
the presence of a stronger relationship. For example, previous work has used reciprocal mentions between accounts to represent dialogue
\cite{ICWSM124575}. An \emph{interactions network} was created with $n$ nodes representing accounts, and $m$ undirected weighted edges
representing reciprocal mentions and retweets between pairs of accounts, with weights corresponding to the number of interactions. All
observed interactions found in the retrieved data sets were considered. Any connected components of size $<5$ were filtered. In addition, we
were also interested in international follower relationships, and a similar undirected unweighted network was created to capture reciprocal
follower links between the accounts in different country sets. Throughout this analysis, due to the sensitivity of the subject matter, and
in the interest of privacy, individual accounts are not identified; instead, we restrict discussion to known extreme right groups and their
affiliates.

\begin{figure}[!b]
	\begin{center}
		\includegraphics[width=0.94\linewidth]{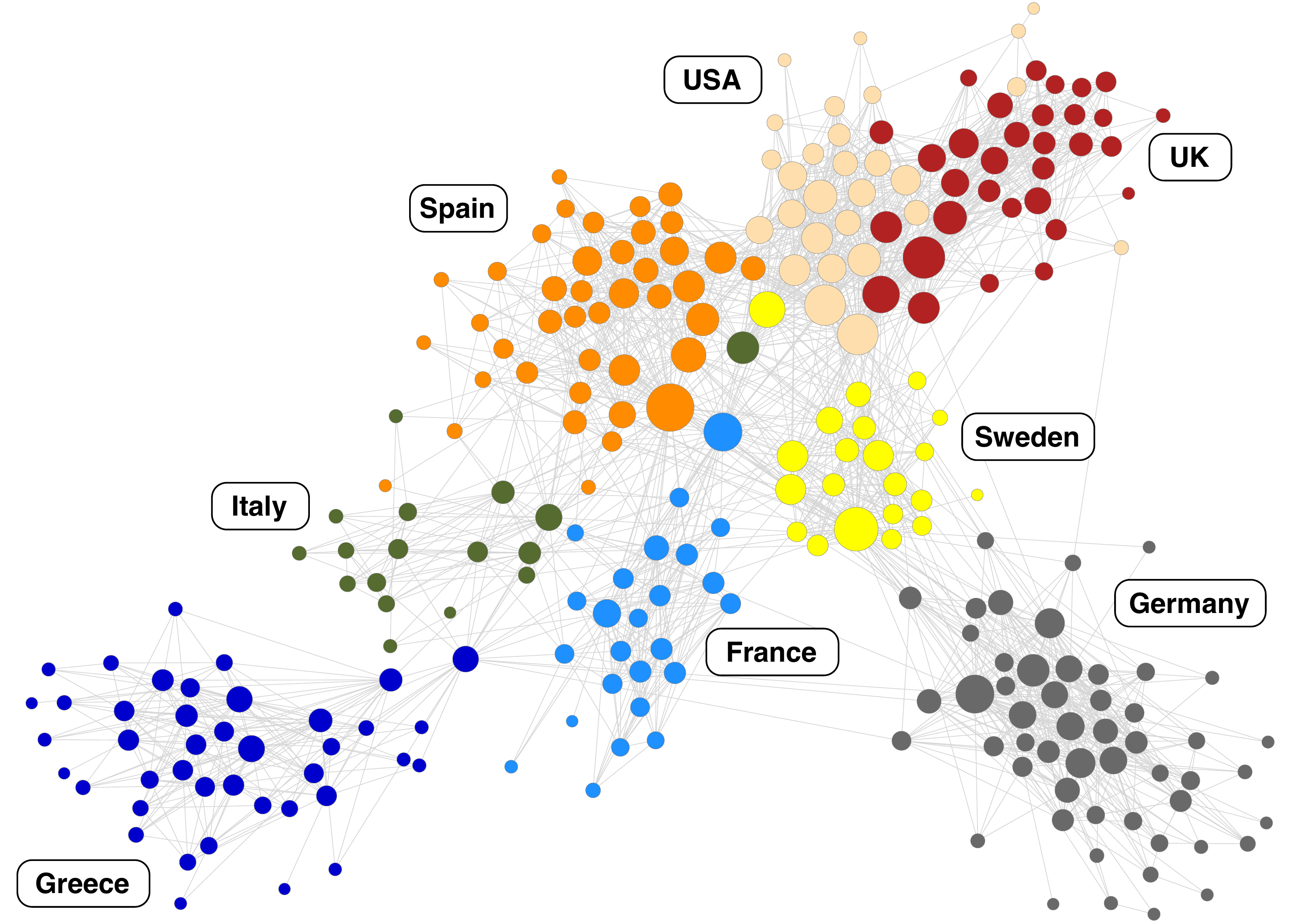}
	\end{center}
	\vskip -1.4em
	\caption{International reciprocal followers network, containing 257 nodes and 2,100 edges. Node size is proportional to degree.}
	\label{fig:globalfollowers}
\end{figure}
 
\subsection{International Follower Awareness}

The international followers network can be seen in Fig.~\ref{fig:globalfollowers}. As might be expected, most of the follower relationships
are between accounts from the same country, although a certain number of international relationships are evident. It would appear that
linguistic and geographical proximity is influential here, for example, we can observe relationships between the Spanish and Italian (and to
a lesser extent, French) accounts, with strong connections also between the UK and USA. Similar behaviour with respect to social ties in
Twitter has been identified by Takhteyev~\etal and Kulshrestha~\etal~\cite{Takhteyev201273,ICWSM124685}. However, there appear to be some
exceptions to the influence of geographical proximity, most notably, Swedish (yellow) and Italian (green) accounts that are not co-located
with their respective country nodes. In both cases, the majority of tweets from these accounts are in English, which presumably ensures a
wider audience. The former account is a Swedish representative of a pan-Scandinavian group espousing national socialist ideas, who appears
to be interacting with many international accounts, particularly from the USA. The Italian account is a national socialist whose tweets
often contain URLs to music or video content hosted on external websites, but it is unclear if a direct association exists with any
particular group. From an analysis of other central nodes in the network (using betweenness centrality), it would seem that those involved
in the dissemination of material via external URLs, or media platforms such as extreme right news websites and radio stations, are
attempting to raise awareness amongst a variety of international followers. This is particularly the case when the English language is used.

\subsection{International Interactions}

\begin{figure}[!b]
	\begin{center}
		\includegraphics[width=0.98\linewidth]{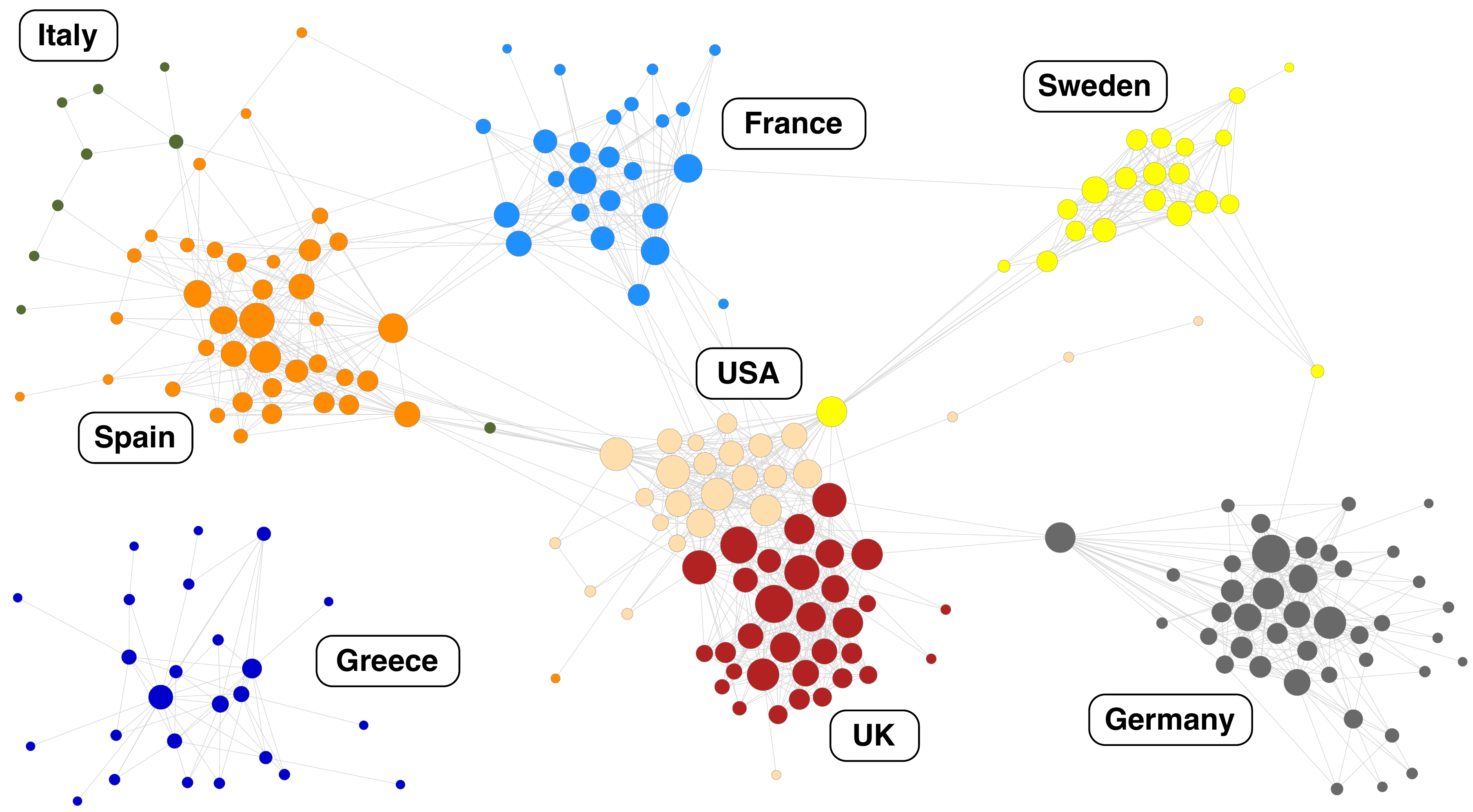}
	\end{center}
	\vskip -1.4em
	\caption{International reciprocal interactions network, containing 218 nodes and 1,186 edges. Node size is proportional to degree.}
	\label{fig:globalmentions}
\end{figure}

The follower-based relationship between international accounts could be considered as passive when compared with that of the interaction
networks, where such interactions can be indicative of actual dialogue between accounts. The interactions network in
Fig.~\ref{fig:globalmentions} is somewhat smaller than the corresponding network in Fig.~\ref{fig:globalfollowers}. We also observe that the
Greek community is now disconnected from the largest connected component. Apart from this, the network has a similar structure to that of
the followers network, in that most interaction occurs within individual country-based communities. Connections between these communities do
exist, but are fewer than in the followers network. The influence of linguistic proximity appears to take precedence here, with the use of
English playing a major role as mentioned in the previous section. For example, a relatively large number of connections remain present
between accounts in the UK and USA. In the case of the German community, while the followers network contains a variety of connections with
other international accounts, this has now been reduced to connections between two German accounts and a small number of UK accounts, in
addition to an account acting as an English language Twitter channel for a Swedish nationalist group. Similarly, the Swedish account
co-located with the USA community is the same account as that in the followers network, who appears to be involved in many English-based
interactions with international accounts.

\section{Local Analysis}
\label{local}

Following the analysis of international relationships described in the previous section, we then proceeded to analyze interactions within
local networks with the objective of detecting specific communities of related accounts, focusing on two data sets as case studies. Based on
our observation of linguistic proximity at the international level, we merged the UK and USA data sets to produce a single English language
data set for one case study, while the second focused on the German language data set. In both cases, we created \emph{expanded} versions of
the initial data sets to facilitate a more detailed analysis, where all available data were also retrieved for those accounts having a
reciprocal follower relationship with more than one of the original identified accounts. As Twitter follower relationships tend to exhibit
lower reciprocity than other social networking sites \cite{Kwak:2010:TSN:1772690.1772751}, the understanding was that this action would be
largely isolated to accounts having a relatively stronger relationship with those from the original data sets. This process resulted in the
inclusion of 1,513 and 448 accounts respectively in the expanded English and German language data sets.

\subsection{Community Detection in Interaction Networks}

For the purpose of community detection based on account interactions, we constructed undirected interaction networks ($G$) from the expanded
data sets. As with the international analysis, only reciprocal edges were used in order to capture stronger relationships, all observed
interactions were included, and connected components of size $<5$ were filtered. We used our variant of the work by Lancichinetti \&
Fortunato to generate a set of stable \textit{consensus communities} from such a network \cite{lanc12consensus,greene12userlists}, where 100
runs of the OSLOM algorithm \cite{lancichinetti11oslom} were used to generate the consensus communities. Following this, the consensus
communities were ranked based on the stability of their members with respect to the corresponding consensus matrix $M$. For a given
consensus community $C$ of size $c$, we computed the mean of the values $M_{xy}$ for all unique pairs ($L_x$, $L_y$) assigned to $C$; this
value has the range $[0, 1]$. We then computed the mean expected value for a community of size $c$ as follows: randomly select $c$ unique
nodes from $G$, and compute their mean pairwise score from the corresponding entries in $M$. This process was repeated over a large number
of randomised runs, yielding an approximation of the expected stability value. The widely-used adjustment technique introduced by
\cite{hubert85compare} was then employed to correct for chance agreement:

\begin{equation}
\textrm{CorrectedStability(\textit{C})} = \frac{\textrm{Stability(\textit{C})} - \textrm{ExpectedStability(\textit{C})}}{1 -
\textrm{ExpectedStability(\textit{C})}}
\label{eqn:stability}
\end{equation}
A value close to 1 will indicate that \textit{C} is a highly-stable community. As higher values of the threshold parameter $\tau$ used with
the consensus method resulted in sparser consensus networks, and having tested with values of $\tau$ in the range $[0.1,0.9]$, we selected
$\tau = 0.5$ as a compromise between node retention and more stable communities. The resolution of communities found by OSLOM is directly
controlled by the associated parameter $P$, which has a default value of $0.5$ in the implementation. Although Lancichinetti~\etal state
that $P=0.1$  delivered an excellent performance on the benchmark graphs used in the OSLOM paper, they also suggest that it would be more
appropriate to estimate $P$ on a case by case basis~\cite{lancichinetti11oslom}. We found that using $P=0.1$ tended to detect a larger
number of relatively small communities. Increasing values of $P$ ($[0.1,0.9]$) produced smaller numbers of larger communities having higher
stability scores (with $\tau = 0.5$). Given this, for both data sets, we used the corresponding value of $P$ that produced the highest mean
stability score to generate communities for detailed analysis. Further details of the $P$ values used can be found later in the case study
sections. For the purpose of this analysis, we focus on larger consensus communities having $\ge10$ members.

\subsection{Community Descriptions}

The content of tweets was also analyzed for the purpose of (1) generating interpretable descriptions for the detected communities, and (2)
identifying latent topics associated with interactions between the accounts assigned to these communities. Following the approach of
Hannon~\etal~\cite{hannon10twitter}, we generated a ``profile document'' for each node in an interactions network, consisting of an
aggregation of their corresponding tweets, from which a tokenized representation was produced. Our initial experiments used all available
terms in the account documents, where the tokenization process involved the exclusion of URLs and stopwords (additional social networking
stopwords such as ``ff'',  ``facebook'' etc. were included with multiple language stopword lists), normalization of diacritics, and stemming
of the remaining terms. Low-frequency terms (appearing in $<4$ account documents) and documents containing $<10$ terms were excluded. These
profile documents were then represented by log-based TF-IDF term vectors, which were subsequently normalized to unit length.

However, we encountered two issues with the use of all available tweet terms as candidates for this representation. It transpired that the
expansion of this data set through the addition of reciprocal follower accounts of the original accounts resulted in the inclusion of
accounts whose mother tongue was not English, for example, accounts from South Africa and Sweden. As a result, generic non-English terms
were treated as highly discriminating due to their relative low frequency within the full set of English terms across all documents.
Separately, the use of all terms required extensive maintenance of a multilingual stopword list. To address these issues, we excluded all
terms apart from hashtags when generating the account documents, which greatly reduced the number of required stopwords, while also
promoting more discriminating non-English terms in subsequent analysis. As it was possible to generate hashtag-based account documents for
$96\%$ of English language accounts and $92\%$ of German language accounts present in the corresponding interaction networks, it was felt
that sufficient coverage of the accounts was retained. In addition, the dimensionality of the vector representations was also considerably
reduced by the sole use of hashtags. Having produced these TF-IDF vectors, each community description was generated by selecting the subset
of vectors for the accounts assigned to the community, and calculating a mean vector $D$ from this subset matrix. The final community
description consisted of the top ten hashtags from $D$. For the remainder of this paper, all hashtags are presented without the preceding
``\#'' character.

\subsection{Topic Analysis}
Topic modelling is concerned with the discovery of latent semantic structure or topics within a set of documents, which can be derived from
co-occurrences of words and documents~\cite{steyvers2006probabilistic}. This strategy dates back to the early work on latent semantic
indexing by Deerwester~\etal~\cite{deerwester90lsi}. Popular methods include probabilistic models such as latent Dirichlet allocation (LDA)
\cite{Blei:2003:LDA:944919.944937}, or matrix factorization techniques such as Non-negative matrix factorization (NMF)~\cite{lee99}. These
have previously been successfully applied in social media analytics, for example, the work of Ramage \etal~\cite{RamageEtAl:10},
Weng~\etal~\cite{Weng:2010:TFT:1718487.1718520}, and Saha and Sindhwani~\cite{Saha:2012:LEE:2124295.2124376}. We initially evaluated both
LDA and NMF-based methods with the account document representations described above, where NMF was found to produce the most
readily-interpretable results. This appeared to be due to the tendency of LDA to discover topics that
over-generalized~\cite{ChemuduguntaGeneralSpecificTopic2006}.

The observed connections between the countries at the international level suggested that specific topics associated with smaller groups of
accounts were present in the extended English and German language data sets. Given this, in the trade-off between generality and
specificity, we opted for the latter in this particular analysis, and so used NMF with the same TF-IDF account vectors as before for topic
analysis. Here, the IDF component ensured a lower ranking for less discriminating terms, thus leading to the discovery of more specific
topics. Having constructed an $n \times m$ term-document matrix $V$, where each column contained a TF-IDF account vector, NMF produced two
factors; $W$, an $n \times T$ matrix containing topic basis vectors and $H$, a $T \times m$ matrix containing the topic assignments for
each account document. To address the instability introduced by random initialization in standard NMF, we employed the NNDSVD method
proposed by Boutsidis and Gallopoulos~\cite{Boutsidis:2008:SBI:1324613.1324653}, which is particularly suitable for sparse matrices.

Although the interactions network topology and general topic analysis provided two different views on a particular data set, in this
analysis, we were primarily interested in the communities within the interactions networks, where the associated topics could be used for
further interpretation of the account membership. As the same hashtag-based account document TF-IDF vectors were used for both community
description generation and topic detection, it was possible to generate a mapping between the detected communities and topics. For each
community, we ranked the topics according to the cosine similarity between its mean community hashtag description vector $D$ and the topic
basis vectors $W$. This method occasionally detected multiple similar topics for a particular community, which was to be expected given that
individual account documents could themselves be associated with multiple topics. To avoid redundant mappings, we ignored any topics having
a cosine similarity $< 0.1$ with each $D$ vector, as using this threshold appeared to produce relevant community-topic mappings in general
for both data sets.

\subsection{Case Study: English language}

\begin{figure}
	\begin{center}
		\includegraphics[width=0.97\linewidth]{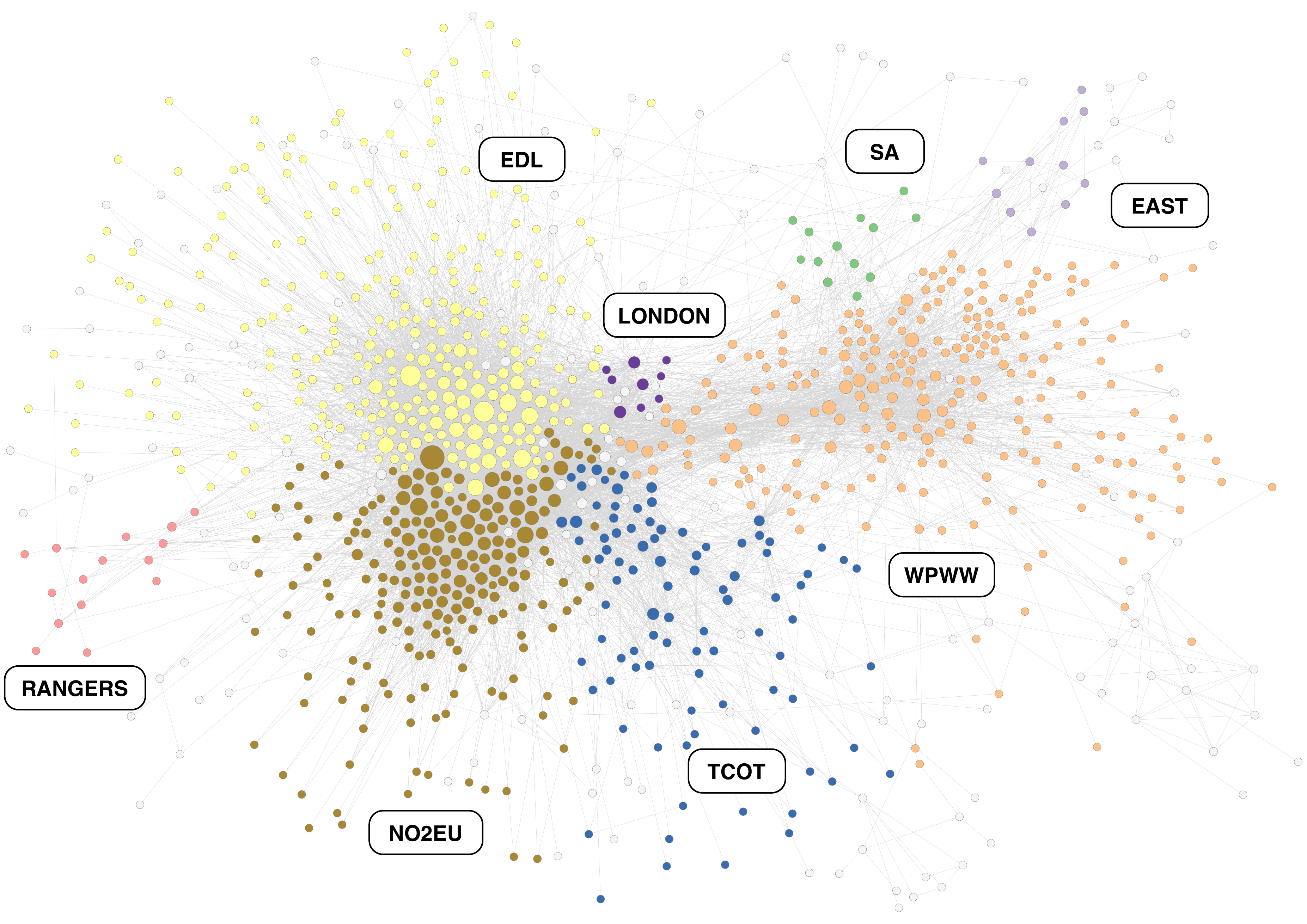}
	\end{center}
	\vskip -2.0em
	\caption{English language interactions network with 1,034 nodes and 9,429 edges. Node size is proportional to degree. Consensus communities
	($P=0.4$, \#members $\geq10$) are labelled.}
	\label{fig:eir}
\end{figure}

\begin{figure}[!t]
	\begin{center}
		\hskip -0.5em
			\includegraphics[scale=0.85]{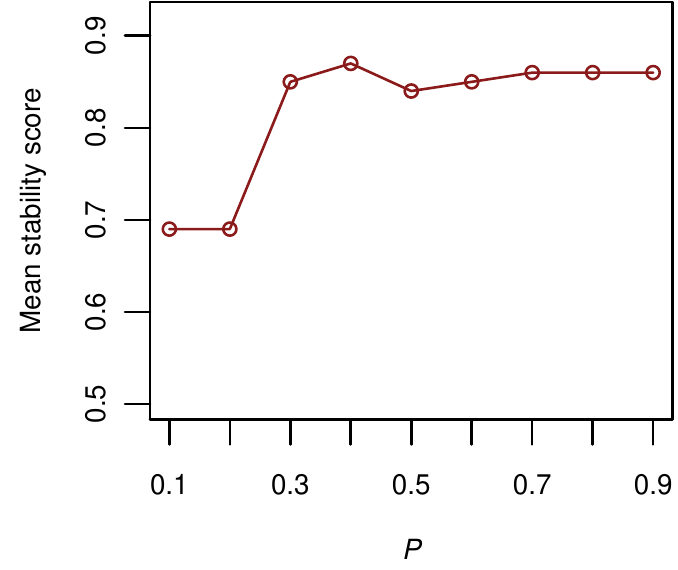}
			\quad
			\includegraphics[scale=0.85]{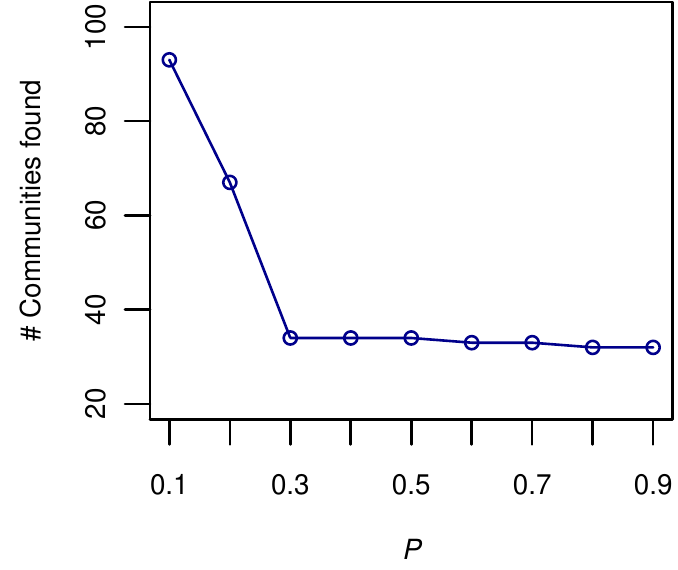}

	\end{center}
	\vskip -1.6em
	\caption{English language mean stability scores (left) and number of communities found (right) for $P$ in $[0.1,0.9]$.}
	\label{fig:engres}
\end{figure}

An interactions network (Fig.~\ref{fig:eir}) was created for the English language data set, consisting of 1,034 nodes and 9,429 edges. Due
to the effect of the resolution parameter $P$ on the communities found by OSLOM, we measured the mean stability score ($\tau=0.5$) and
number of communities found for values of $P$ in $[0.1,0.9]$, using the consensus community method previously described. A plot of the mean score
and sizes can be found in Fig.~\ref{fig:engres}, where it can be seen that the highest mean score was generated for $P=0.4$, while the
number of communities found had also stabilized at $\sim34$. Eight communities having at least ten members were found for $P=0.4$ ($86\%$ of
total network account nodes), and their corresponding hashtag descriptions and stability scores can be found in
Table~\ref{tab:engcommsres04}. These communities have been manually annotated with identifiers for reference in the subsequent discussion,
based on an analysis of the account member composition (for example, EAST, LONDON). Topic analysis was also performed using NMF, with number
of topics $T=15$. Having experimented with a range of values for $T$, we used $T=15$ as this was the smallest value which led to the
emergence of non-English language topics.

\def\imagetop#1{\vtop{\null\hbox{#1}}}

\begin{table}[!b]
\caption{Consensus communities found in the English language interactions network ($\tau=0.5$, $P=0.4$, \#members $\geq10$), representing
$86\%$ of all network nodes.}
\begin{center}
\vskip -0.4em
\begin{tabular}{| p{1.67cm} | p{7.95cm} | r | c |}
\hline 
\emph{Id} & \emph{Description} & \hspace{0.1mm} \emph{Size} \hspace{0.1mm}  & \hspace{0.1mm} \emph{Score} \hspace{0.1mm}\\ \hline \hline
SA  &
{\small
\textsf{anc, onswiloorleef, southafrica, ancyl, onssaloorleef, zuma, stopabsa, zumaspear, nuus, svpol}}
& 13 & 1.00 \\
\hline
EAST  &
{\small
\textsf{presstv, sv, wpww, metal, estonia, 666, polska, ww2, edl, ukraine}}
& 11 & 0.95 \\ \hline
WPWW  &
{\small
\textsf{wpww, tcot, whitepower, nigger, p2, niggers, teaparty, gop, obama, whitepride}}
& 240 & 0.88 \\ \hline
TCOT  &
{\small
\textsf{tcot, teaparty, p2, obama, gop, israel, tlot, lnyhbt, sgp, islam}}
& 91 &  0.74 \\ \hline
NO2EU  &
{\small
\textsf{no2eu, ukip, bbcqt, labour, edl, leveson, newsnight, euro2012, eurovision, london2012}}
& 214 & 0.74 \\ \hline
EDL  &
{\small
\textsf{edl, uaf, islam, bbcqt, lfc, bnp, muslim, tcot, mufc, israel}}
& 290 & 0.62 \\ \hline
RANGERS  &
{\small
\textsf{watp, rangersfamily, nosurrender, gersfollowback, wedontdowalkingaway, rfc, taintedtitle, rangersfamilly, rangers, rangerfamily}}
& 16 & 0.58 \\ \hline
LONDON  &
{\small
\textsf{edl, coys, stgeorgesday, londonriots, whys, savages, bbcqt, stfc, tottenham, dench}}
& 10 & 0.53 \\ \hline
\end{tabular}
\end{center}
\vskip -2.4em
\label{tab:engcommsres04}
\end{table}

The accounts in the SA community appear to be white South Africans, with some profiles containing racist and national socialist references.
Many tweets from these accounts relate to perceived cultural threats from black South Africans that are often retweeted by international
accounts. Most of the description hashtags are related to South Africa, such as \emph{anc} (\textit{African National Congress}, the current
governing party) and \emph{onswiloorleef} (``We want to survive"), associated with a campaign highlighting alleged violent attacks against
Afrikaners. This community is associated with a single South African topic, which is to be expected given the discriminating nature of
these hashtags. The EAST community includes white power/national socialist accounts from Eastern European countries such as Estonia, Poland
and Ukraine, who occasionally tweet in English. Notable description hashtags include \emph{wpww} (white pride world wide) and \emph{metal},
where the latter refers to a music sub-genre known as nationalist socialist black metal. Of similar interest is \emph{presstv}, the Iranian
state-owned English language news network. This has previously been accused of propagating anti-Semitic content, while also hosting
Holocaust-deniers and white nationalists. The most similar topic for this community is one connected to white pride ideology that includes
media references, for example, \emph{wpradio} (white pride radio).

\begin{table}[b!]
\caption{English language communities and associated NMF topics ($T=15$, hashtag description cosine similarity $\geq0.1$).}
\begin{center}
\vskip -0.4em
\begin{tabular}{| l |c | p{8.7cm} |}
\hline 
\emph{Community} & \emph{Similarity} & \emph{Top 10 Topic Terms} \\ \hline \hline
SA & 0.56 &
{\small
\textsf{anc, onswiloorleef, stopabsa, ancyl, southafrica, zuma, nuus, afrikaans, zumaspear, genocide}}
\\ \hline \hline 
EAST & 0.12 &
{\small
\textsf{wpww, whitepride, wpradio, contest, staywhite, whiteisright, white, whiteunity, genocide, skinhead}}
\\ \hline \hline
\multirow{3}{*}{WPWW} 
& 0.14 &
{\small
\textsf{wpww, whitepride, wpradio, contest, staywhite, whiteisright, white, whiteunity, genocide, skinhead}}
 \\ \cline{2-3}
& 0.10 &
{\small
\textsf{tcot, teaparty, p2, gop, tlot, obama, sgp, ocra, lnyhbt, twisters}}
 \\ \hline \hline
\multirow{5}{*}{TCOT}
 & 0.48 &
{\small
\textsf{tcot, teaparty, p2, gop, tlot, obama, sgp, ocra, lnyhbt, twisters}}
\\ \cline{2-3}
 & 0.17 &
{\small
\textsf{israel, islam, sharia, muslim, jihad, iran, gaza, syria, egypt, nadarkhani}}
\\ \cline{2-3}
 & 0.15 &
{\small
\textsf{prolife, abortion, tcot, prochoice, personhood, god, octoberbaby, 912, gingrich, preborn}}
\\ \hline \hline
\multirow{8}{*}{NO2EU} & 0.16 &
{\small
\textsf{labour, leveson, occupylsx, cameron, ukuncut, bnp, greece, bbc, syria, olympics}}
\\ \cline{2-3}
 & 0.13 &
{\small
\textsf{bbcqt, newsnight, eurovision, london2012, euro2012, closingceremony, pmqs, teamgb, leveson, olympics}}
\\ \cline{2-3}
 & 0.12 &
{\small
\textsf{no2eu, lab11, english, labour, england, eurozone, euro, eurocrash, euro2012, london2012}}
\\ \cline{2-3}
 & 0.12 &
{\small
\textsf{ukip, voteukip, christappin, uk, greece, ronpaul, freegary, tories, richardo, extradition}}
\\ \hline \hline
EDL & 0.42 &
{\small
\textsf{edl, uaf, islam, bnp, casualsunited, luton, rochdale, mdl, bristol, praetorian}}
\\ \hline \hline 
RANGERS & 0.59 &
{\small
\textsf{rangersfamily, watp, nosurrender, rangers, gersfollowback, rfc, wedontdowalkingaway, taintedtitle, rangersfamilly,
celtictaintedtitle}} \\ \hline \hline 
LONDON & 0.15 &
{\small
\textsf{edl, uaf, islam, bnp, casualsunited, luton, rochdale, mdl, bristol, praetorian}}
\\ \hline
\end{tabular}
\end{center}
\vskip -2.8em
\label{tab:engcommtopics}
\end{table}

The WPWW community would appear to be national socialist/white power in nature, with the appearance of hashtags such as \emph{wpww} and
\emph{whitepower}. An analysis of the accounts and associated profiles finds references to the \textit{American Nazi Party}, along with
other related terms such as \textit{14} (a reference to a 14-word slogan coined by the white supremacist David Lane), and \textit{88}
(``heil hitler'') in account names. There are also references to skinhead groups, including a website where related media and merchandise
can be found for sale. Accounts appear to be mostly associated with the USA, although a small number of European accounts are also present.
References to \emph{tcot} (top conservatives on Twitter) and \emph{gop} (US Republican Party) can also be seen, indicating the presence of
more traditional conservative accounts. However, this does not necessarily point to any official link between these groups. Two topics are
most closely associated with this community, mirroring the account and description hashtag division between \emph{wpww} and \emph{tcot}. The
TCOT community is largely composed of traditional conservatives, where the description contains a number of hashtags commonly used by these
groups such as \emph{gop} and \emph{teaparty}. However, we also note the presence of \emph{p2} (progressives, Tweet Progress), effectively
the polar opposite of \emph{tcot}. A number of anti-Islamic counter-Jihad accounts are also present~\cite{GoodwinEDLCounterJihad2013}. This
community is strongly linked to the \emph{tcot} topic, with the counter-Jihad and pro-life/anti-abortion topics also of interest.

Opposition to the EU appears to be the binding theme of the NO2EU community, which contains several political or electoral accounts,
including a number affiliated with British Eurosceptic parties such as the \emph{United Kingdom Independence Party} (UKIP). Non-electoral
British nationalist accounts are also present, where their tweets and profiles often contain anti-EU statements and imagery. We also see
references to British media such as BBC current affairs programmes (\emph{bbcqt}, \emph{newsnight}). Accordingly, topics linked to this
community appear to be concerned with politics and the EU in general. The EDL community consists mostly of accounts associated with the
\textit{English Defence League} (\emph{edl}), a counter-Jihad movement opposed to the alleged spread of radical Islamism within the
UK~\cite{GoodwinEDLCounterJihad2013,BartlettInsidedEDL2011}. Other accounts include those associated with \emph{Casuals United}, a protest
group linked with the EDL that formed from an alliance of football hooligans (references to football clubs can also be observed). The
\emph{uaf} hashtag refers to the \textit{Unite Against Fascism} group; a staunch opponent of the EDL. Accounts from the USA are also
present, acting as a further reminder of the international relationships within the counter-Jihad
movement~\cite{GoodwinEDLCounterJihad2013}. The final two small communities appear to consist of soccer fans, and are associated with
Rangers Football Club from Scotland, and London-based soccer clubs respectively. The latter also contains a number of accounts affiliated
with the EDL.

\subsection{Case Study: German language}

\begin{figure}[!t]
	\begin{center}
		\includegraphics[width=0.97\linewidth]{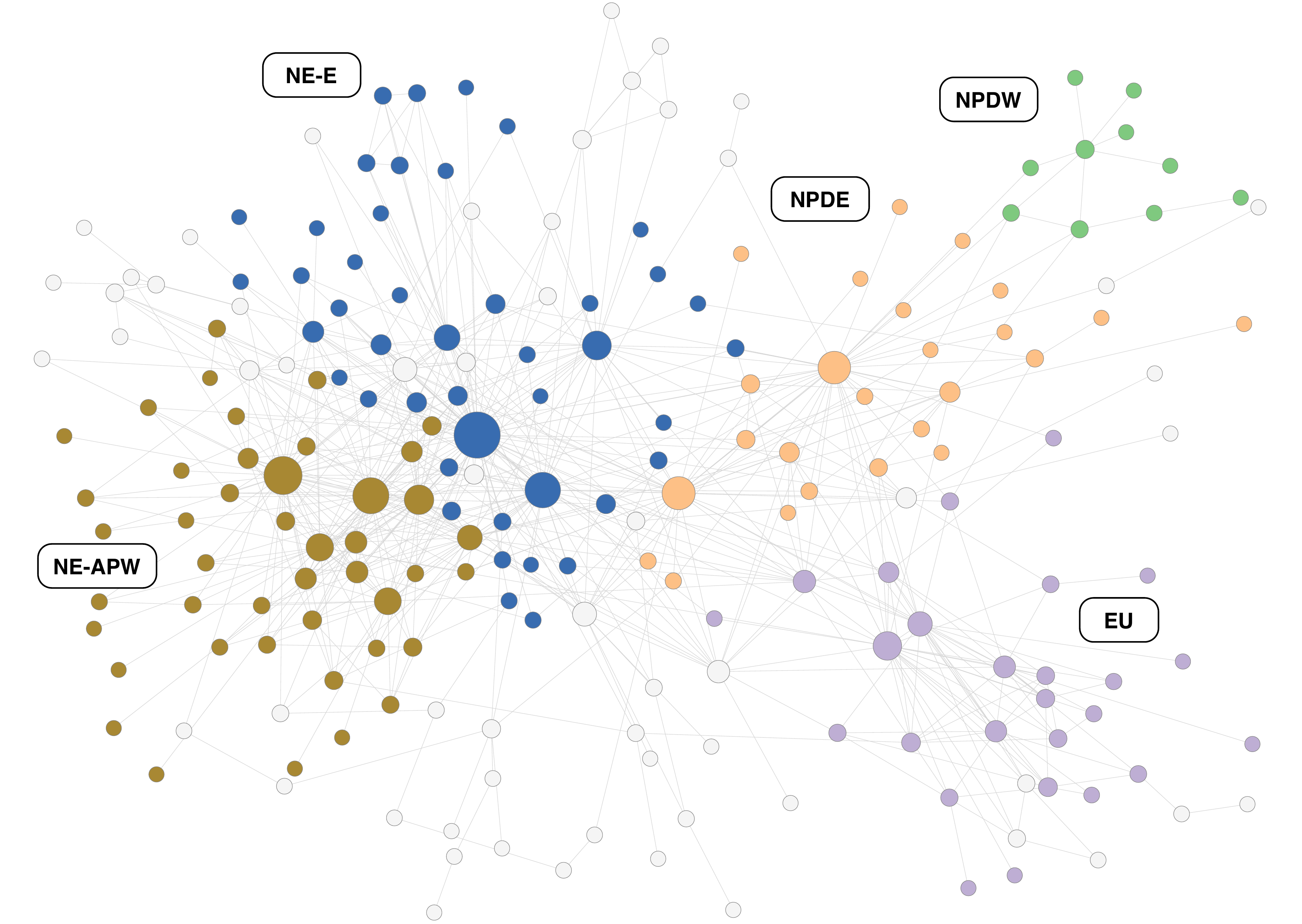}	\end{center}
	\vskip -1.9em
	\caption{German language interactions network with 208 nodes and 630 edges. Node size is proportional to degree. Consensus communities
	($P=0.9$, \#members $\geq10$) are labelled. }
	\label{fig:gir}
\end{figure}

\begin{figure}[b!]
	\begin{center}
	\hskip -0.6em
			\includegraphics[scale=0.85]{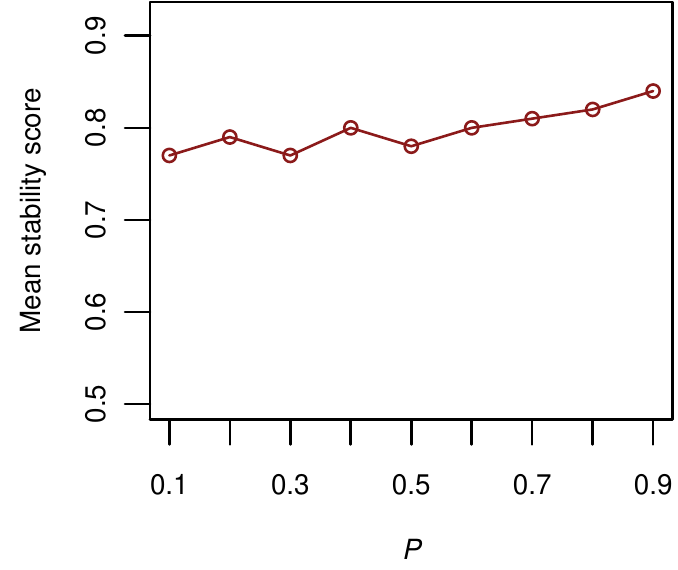}
			\quad
			\includegraphics[scale=0.85]{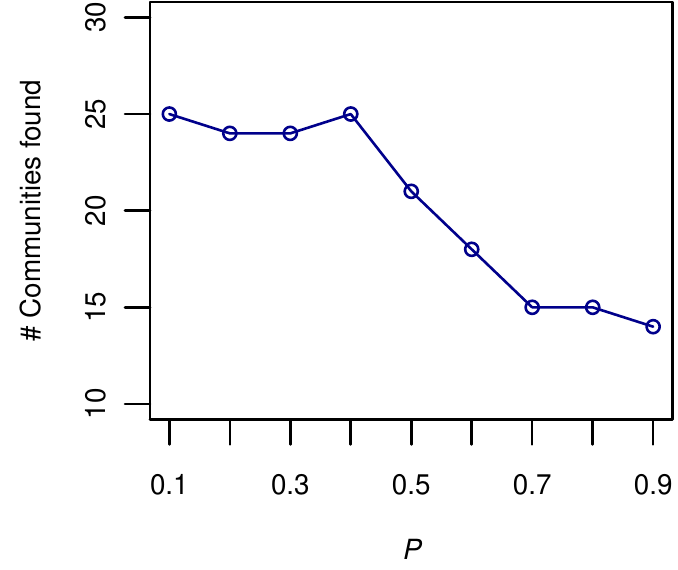}
	\end{center}
	\vskip -1.8em
	\caption{German language mean stability scores (left) and number of communities found (right) for $P$ in $[0.1,0.9]$.}
	\label{fig:gerres}
\end{figure}

An interactions network (Fig.~\ref{fig:gir}) was created for the German language data set, consisting of 208 nodes and 630 edges. As before,
we measured the mean stability score ($\tau=0.5$) and number of communities found for values of $P$ in $[0.1,0.9]$. The corresponding plots
found in Fig.~\ref{fig:gerres} show that the highest mean score was generated for $P=0.9$, while the number of communities found had also
stabilized at $\sim15$. Five communities having at least ten members were found for $P=0.9$ ($74\%$ of total network account nodes), and
their corresponding hashtag descriptions, stability scores and manually annotated identifiers can be found in Table~\ref{tab:gercommsres09}.
Topic analysis was also performed with NMF, with number of topics $T=10$, as lower values of $T$ led to topics associated with relatively
few account documents.

\begin{table}[b!]
\caption{Consensus communities found in the German language interactions network ($\tau=0.5$, $P=0.9$, \#members $\geq10$), representing
$74\%$ of all network nodes.}
\begin{center}
\vskip -0.4em
\begin{tabular}{| p{1.67cm} | p{7.95cm} | r | c |}
\hline 
\emph{Id} & \emph{Description} & \hspace{0.1mm} \emph{Size} \hspace{0.1mm}  & \hspace{0.1mm} \emph{Score} \hspace{0.1mm}\\ 
\hline \hline
NPDW  &
{\small
\textsf{npd, bochum, nrw, landtagswahl, wattenscheid, ovg, flugblatt, krefeld, anklage, bamberg}}
& 10 & 0.92 \\ \hline
EU &
{\small
\textsf{euro, esm, stopesm, spd, islam, cdu, piraten, fdp, berlin, griechenland}}
& 26 & 0.82 \\ \hline
NPDE &
{\small
\textsf{npd, ltwlsa, deutschland, ltw2011, berlin, linke, lsa, ltwmv, guttenberg, sarrazin}}
& 28 & 0.80 \\ \hline
NE-E  &
{\small
\textsf{geithain, apw, emwall, widerstand, jena, unsterblichen, gera, leipzig, volkstod, tdi}}
& 44 &  0.71 \\ \hline
NE-APW  &
{\small
\textsf{apw, gema, denkdran, volkstod, demokraten, dresden, hannover, spreelichter, 130abschaffen, unsterblichen}}
& 45 & 0.64 \\ \hline
\end{tabular}
\end{center}
\vskip -2.8em
\label{tab:gercommsres09}
\end{table}

The first community consists of accounts associated with the \textit{Nationaldemokratische Partei Deutschlands - National Democratic Party
of Germany} (NPD) that appear to be localized to western regions of Germany. Hashtags such as \emph{landtagswahl/ltw\ldots} (regional
election) and \emph{flugblatt} (flyer/leaflet/pamphlet) along with analysis of the tweet content may indicate mobilization prior to
elections. Two topics most closely associated with this community include a Germany-wide NPD topic in addition to a general topic appearing
to be related to street demonstrations. For example, \emph{stolberg} refers to the town where a German teenager was killed by non-Germans in
2008, which has been the focus of annual extreme right commemorations. The EU community appears to be somewhat analogous to the NO2EU
English language community, in that its membership is composed of a mixture of moderate and more extreme nationalist accounts bound by
general opposition to the EU and related entities such as the European Stability Mechanism (\emph{stopesm}), in addition to domestic German
political parties (\emph{cdu}, \emph{fdp}, \emph{spd}). Other notable members include a number of counter-Jihad accounts, along with others
associated with relatively high-profile external media and blog websites.

The NPDE community is the second that can be associated with the NPD, which, in contrast to the NPDW community, appears to be largely
localized to eastern German regions. In addition to NPD politicians, various Freies Netz (neo-Nazi collectives - FN) and ``information/news
portal'' accounts are also present~\cite{NGNPortals2012}. Similar variants of the \emph{landtagswahl} hashtag and corresponding tweets to
those of NPDW can also be observed. Other issues of interest include references to Thilo Sarrazin, a German politician and former Bundesbank
executive who has criticized German immigration policy and proposed the abolition of the euro currency. As might be expected, the general
NPD topic is linked to this community, while the anti-EU/political system topic is also prominent.

\begin{table}[b!]
\caption{German language communities and associated NMF topics ($T=10$, hashtag description cosine similarity $\geq0.1$).}
\begin{center}
\vskip -0.4em
\begin{tabular}{| l | c | p{8.7cm} |}
\hline 
\emph{Community} & \emph{Similarity} & \emph{Top 10 Topic Terms} \\ \hline \hline
\multirow{4}{*}{NPDW} 
& 0.35 &
{\small
\textsf{npd, landtagswahl, ltwlsa, ltwmv, nrw, sachsenanhalt, bochum, linke, wahlen, wattenscheid}}
 \\ \cline{2-3}
& 0.14 &
{\small
\textsf{stolberg, dortmund, nrw, demonstration, demo, aachen, koln, brd, rheinland, munster}}
 \\
\hline \hline
EU & 0.39 &
{\small
\textsf{esm, euro, stopesm, spd, cdu, deutschland, fdp, piraten, stoppesm, islam}}
\\ \hline \hline
\multirow{4}{*}{NPDE}
 & 0.26 &
{\small
\textsf{npd, landtagswahl, ltwlsa, ltwmv, nrw, sachsenanhalt, bochum, linke, wahlen, wattenscheid}}
\\ \cline{2-3}
 & 0.14 &
{\small
\textsf{esm, euro, stopesm, spd, cdu, deutschland, fdp, piraten, stoppesm, islam}}
\\ \hline \hline
\multirow{10}{*}{NE-E} & 0.17 &
{\small
\textsf{emwall, geithain, tdi, leipzig, imc, linksunten, arbeiterkampftag, lvz, hof, dd2012}}
\\ \cline{2-3}
 & 0.16 &
{\small
\textsf{widerstand, unsterblichen, altermedia, israel, wuppertal, hannover, dieunsterblichen, repression, abschiebar, spreelichter}}
\\ \cline{2-3}
 & 0.15 &
{\small
\textsf{jena, raz10, gera, thuringen, volkstod, kahla, altenburg, demokraten, dresden, apw}}
\\ \cline{2-3}
 & 0.13 &
{\small
\textsf{apw, volkstod, hannover, heldengedenken, demokratie, spreelichter, cottbus, guttenberg, dresden, vds}}
\\ \cline{2-3}
 & 0.10 &
{\small
\textsf{stolberg, dortmund, nrw, demonstration, demo, aachen, koln, brd, rheinland, munster}}
\\ \hline \hline
\multirow{8}{*}{NE-APW} & 0.40 &
{\small
\textsf{apw, volkstod, hannover, heldengedenken, demokratie, spreelichter, cottbus, guttenberg, dresden, vds}}
\\ \cline{2-3}
 & 0.19 &
{\small
\textsf{gema, denkdran, chemnitz, 5maerz, dresden, magdeburg, 13februar, 130abschaffen, demokraten, akt}}
\\ \cline{2-3}
 & 0.14 &
{\small
\textsf{widerstand, unsterblichen, altermedia, israel, wuppertal, hannover, dieunsterblichen, repression, abschiebar, spreelichter}}
\\ \cline{2-3}
 & 0.13 &
{\small
\textsf{jena, raz10, gera, thuringen, volkstod, kahla, altenburg, demokraten, dresden, apw}}
\\ \hline
\end{tabular}
\end{center}
\vskip -2.8em
\label{tab:gercommtopics}
\end{table}

The remaining two communities contain accounts associated with a variety of non-electoral groups and individuals within the German extreme
right. An analysis of the accounts in the NE-E community finds them to be associated with eastern regions of Germany, in particular, the
town of Geithain near Leipzig in Sachsen. Accounts assocatiated with various extreme right groups are present, such as FN and the
\textit{Junge Nationaldemokraten} (Young National Democrats, youth wing of the NPD). References to \textit{Aktionsb\"{u}ros} (coordination
of activist activities) are also made. The \emph{emwall} hashtag was popularly used during the 2012 UEFA European Football Championship, and
accounts in this community used it to promote tweets suggesting that players having non-German ancestry should be excluded from the national
squad. Separately, the \emph{unsterblichen} (immortals) hashtag refers to anti-democratic flashmob marches that previously occurred
sporadically throughout Germany in 2011 and 2012. These protests were linked to \textit{Spreelichter}, an organization that was banned by
the German authorities in 2012, whose account was also a member of this community~\cite{ZeitSpreelichterBan2012}. They used social media to
propagate national socialist-related material, including professional-quality videos of the marches themselves. In general, the accounts in
this community appear to be quite active, with many tweets containing URLs linking to content hosted on external platforms such as YouTube
or other dedicated websites.

The second non-electoral community, NE-APW, appears to contain accounts from regions throughout Germany. The concept of \emph{apw} (au\ss
erparlamentarischer Widerstand - non-parliamentary resistance) is prominent here, referring to actions taken outside of the democratic
process. Other relevant hashtags include \emph{13februar}, \emph{denkdran}, \emph{dresden} and \emph{gema}, which all refer to the bombing
of Dresden which began on February 13, 1945. The anniversary of this event is usually commemorated by extreme right groups each year. Also
relevant is \emph{volkstod}, which refers to the perceived destruction of the German race and traditions since World War II. This concept
has separately featured in material distributed by alleged supporters of the National Socialist Underground~\cite{ZeitEmingerNSU2013}, a
group linked to a series of murders throughout Germany. The official account of the \emph{Besseres Hannover} organization is present here;
an initial ban by the German authorities was followed by this account being blocked by Twitter within
Germany~\cite{ZeitBesseresHannoverBan2012,GuardianBesseresHannoverTwitter2012}. Both NE communities are associated with a variety of topics,
perhaps reflecting the fragmented nature of non-electoral groups in the German extreme right. Relevant themes include street demonstrations
(\emph{stolberg}), some form of resistance (\emph{apw}, \emph{widerstand}) and media references such as \emph{altermedia} (a collective of
politically-incorrect/nationalist-oriented news websites).

\subsection{Case Studies Discussion}

For both the English and German language case studies, identifiable communities of accounts and related topics are clearly observable. When
creating the original data sets, we purposely focused on non-electoral accounts and excluded electoral accounts affiliated with political
parties. However, in both cases, the expansion of the data sets using reciprocal follower relationships resulted in the inclusion of
electoral accounts, such as those affiliated with UKIP or the NPD. This potentially indicates the presence of some form of relationship
between these two categories, if only at a passive level. Other similarities include the presence of distinct anti-EU communities and
topics, in addition to the use of media accounts such as those associated with extreme right news websites and radio stations, along with
external websites hosting media content. While geographical proximity is evident in most communities, linguistic proximity is a key factor
in the existence of international connections such as those between certain counter-Jihad groups and individuals. Although the underlying
ideology of certain communities can often be identified, this is less clear in other cases, especially when the mappings between communities
and their associated topics are considered. Here, multiple topic mappings suggest a complexity and diversity in both the membership
composition and interests of the corresponding community. However, this may also be related to data incompleteness, variances in Twitter
usage patterns between different countries, and the fact that an opinion that may be legally voiced in one country may not be permitted in
another.

We should mention that this sample of accounts does not provide full coverage of all extreme right Twitter activity, and the accounts and
subsequent communities and topics are greatly dependent on the initial selection of relevant accounts. In this domain, the random sampling
of Twitter accounts is unlikely to yield a representative data set, as it is probably safe to assume that the total number of extreme right
accounts merely constitutes a small percentage of all accounts. The fact that we did not rely on hashtags for data selection, coupled with
the expansion using reciprocal follower relationships goes some way to avoid ``selecting on the dependent
variable"~\cite{TufekciBigDataPitfalls2013}. However, as suggested by Boyd and Crawford, it is important to acknowledge all known data set
limitations~\cite{doi:10.1080/1369118X.2012.678878}. At the same time, the authors also recognize the value of small data sets, where
research insights can be found at any level. In this case, in spite of data sampling and coverage issues, it is still possible to detect the
presence of extreme right communities and topics on Twitter. In addition, they emphasize the importance of results interpretation, which we
have addressed here in our discussion of the communities and topics. In relation to this, care should be taken when inferring conclusions
from results. It is unclear as to how representative they may be of offline extreme right network activity. It may be the case that social
media platforms are merely used by these groups to disseminate related material to a wider audience, with the majority of subsequent
interaction occurring elsewhere, but it is naturally difficult to quantify the extent to which this occurs. Separately, Boyd and Crawford
also raise the issue of ethics in relation to publicly accessible data, emphasising the need for accountability on the part of researchers.
Here, we address this by restricting discussion to known extreme right groups and their affiliates without identifying any individual
accounts.

\section{Conclusions and Future Work}
\label{conclusions}

Extreme right groups have become increasingly active in social media platforms such as Twitter in recent years. We have presented an
analysis of the activity of a selection of such groups using network representations based on reciprocal follower and interaction activity,
in addition to topic analysis of their corresponding tweets. The existence of stable communities and associated topics within local
interaction networks has been demonstrated, and we have also identified international relationships between groups across geopolitical
boundaries. Although a certain awareness exists between accounts based on follower relationships, it would appear that mentions and retweets
interactions indicate stronger relationships where linguistic and geographical proximity are highly influential, in particular, the use of
the English language. In relation to this, media accounts such as those associated with extreme right news websites and radio stations,
along with external websites hosting content such as music or video, are a popular mechanism for the dissemination of associated material.

In future work, we will address the issues of sampling and incompleteness in the data sets, where the emergence of new extreme right groups
should also be considered. The temporal properties of these networks will also be studied to provide insight into the evolution of extreme
right communities over time. Separately, we plan to investigate  the use of probabilistic topic models that support the discovery of more
specific topics similar to those found in this analysis.

\vspace{3 mm}\noindent\emph{Acknowledgements.}
This research was supported by 2CENTRE, the EU funded Cybercrime Centres of
 Excellence Network and Science Foundation Ireland Grant 08/SRC/I1407 (Clique: Graph and Network Analysis Cluster).
 
\bibliography{20140127_OCallaghan_Extreme_Right_Social_Media}
\bibliographystyle{splncs}

\end{document}